\documentclass[12pt,preprint]{aastex}

\usepackage{lineno}
\linenumbers

\usepackage{amsmath}
\usepackage{graphicx}
\usepackage{url}
\usepackage{multirow}

\usepackage{hyperref}

\usepackage{color}
\usepackage{xspace}

\newcommand{\fermi}{{\textit{Fermi}}\xspace}
\newcommand{\lat}{LAT\xspace}

\newcommand{\pointlike}{\ensuremath{\mathtt{pointlike}}\xspace}
\newcommand{\gtlike}{\ensuremath{\mathtt{gtlike}}\xspace}

\newcommand{\ts}{\ensuremath{\text{TS}}\xspace}
\newcommand{\tsext}{\ensuremath{\ts_\text{ext}}\xspace}
\newcommand{\tsbreak}{\ensuremath{\ts_\text{break}}\xspace}
\newcommand{\ebreak}{\ensuremath{E_\text{break}}\xspace}

\newcommand{\ph}{\text{ph}\xspace}
\newcommand{\erg}{\text{erg}\xspace}
\newcommand{\cm}{\text{cm}\xspace}
\newcommand{\s}{\text{s}\xspace}
\newcommand{\mev}{\text{MeV}\xspace}
\newcommand{\gev}{\text{GeV}\xspace}

\newcommand{\fluxunits}{\ensuremath{\ph\;\cm^{-2}\s^{-1}}\xspace}
\newcommand{\efluxunits}{\ensuremath{\erg\;\cm^{-2}\s^{-1}}\xspace}
\newcommand{\stat}{\text{stat}\xspace}

\newcommand{\secref}[1]{\S~\ref{sec:#1}}

\newcommand{\tabref}[1]{Table~\ref{tab:#1}}
\newcommand{\figref}[1]{Figure~\ref{fig:#1}}

\newcommand{\degree}{\ensuremath{^\circ}\xspace}

\newcommand{\likelihood}{\ensuremath{\mathcal{L}}\xspace}

\begin{document}

\title{\fermi-LAT Detection of a Break in the Gamma-Ray Spectrum of the Supernova Remnant Cassiopeia A}
\shorttitle{\fermi-LAT Observations of Cassiopeia A}


\author{
Y.~Yuan\altaffilmark{1,2},
S.~Funk\altaffilmark{1,3},
G.~J\'ohannesson\altaffilmark{4},
J.~Lande\altaffilmark{1,5},
L.~Tibaldo\altaffilmark{1},
Y.~Uchiyama\altaffilmark{6,7}
}
\altaffiltext{1}{W. W. Hansen Experimental Physics Laboratory, Kavli Institute for Particle Astrophysics and Cosmology, Department of Physics and SLAC National Accelerator Laboratory, Stanford University, Stanford, CA 94305, USA}
\altaffiltext{2}{email: yuanyj@stanford.edu}
\altaffiltext{3}{email: funk@slac.stanford.edu}
\altaffiltext{4}{Science Institute, University of Iceland, IS-107 Reykjavik, Iceland}
\altaffiltext{5}{email: joshualande@gmail.com}
\altaffiltext{6}{3-34-1 Nishi-Ikebukuro, Toshima-ku, Tokyo, Japan 171-8501}
\altaffiltext{7}{email: uchiyama@slac.stanford.edu}

\begin{abstract}
  We report on observations of the supernova remnant Cassiopeia
A in the energy range from 100 MeV to 100 GeV using 44 months of
observations from the Large Area Telescope on board the \textit{Fermi Gamma-ray
Space Telescope}.  We perform a detailed spectral analysis of this source
and report on a low-energy break in the spectrum at $1.72^{+1.35}_{-0.89}$
\gev. By comparing the results with models for the $\gamma$-ray emission,
we find that hadronic emission is preferred for the GeV energy range.

\end{abstract}



\section{Introduction}

With an age of $\sim 350$ years, the supernova remnant (SNR)
Cassiopeia A (Cas A) is one of the youngest objects of this class in
our Galaxy. It is also one of the best studied objects with both
thermal and non-thermal broad-band emission ranging from radio through
X-ray all the way to GeV and TeV gamma rays. It is the brightest radio
source in the sky outside of our solar system \citep{Baars} and is
located at a distance of
3.4$^{+0.3}_{-0.1}$~kpc~\citep{Reed1995}. Non-thermal emission tracing
the acceleration of particles to relativistic energies has been
detected in both the forward and reverse shocks~\citep[see
e.g.\ ][]{Gotthelf2001, Hughes2000, HelderVink2008, Maeda2009}, in
particular seen through high-angular resolution X-ray studies. Fast
variability and small filaments seen in these X-ray observations also
suggest rather large magnetic fields of 0.1-0.3 mG in the shock region of Cas
A~\citep{PatnaudeFesen2007, PatnaudeFesen2009, Uchiyama2008}.  The
observed brightness variations might, however, also be produced by
local enhancements of the turbulent magnetic field~\citep{Bykov2008}.

Gamma-ray observations further corroborate the existence of
non-thermal particles in the shell of Cas A. The SNR was first
detected at TeV energies with the HEGRA telescope
system~\citep{HEGRA:casA}, later confirmed by MAGIC~\citep{MAGIC:casA}
and VERITAS~\citep{VERITAS_CasA}, and subsequently detected at lower
(GeV) energies with the Large Area Telescope (\lat)
on board the \textit{Fermi Gamma-ray Space Telescope} (\fermi)~\citep[Paper~I,][]{PaperI}. Those
observations revealed a rather modest gamma-ray flux, compared to the
synchrotron radio through X-ray emission, further strengthening the
argument for a rather high magnetic field. The field can hardly be
significantly less than 100~$\mu$G~\citep{PaperI}, consistent
with earlier studies~\citep[see e.g.][]{Vink2003, Parizot2006}. It
should be stressed that the magnetic field is likely to be
non-uniform. This was originally proposed by~\citet{Atoyan2000b} who suggested
greatly amplified magnetic fields of up to 1 mG in compact filaments.
Because both the photon and matter densities in the shock regions are
rather high, these gamma-ray studies also suggested that the
non-thermal electron (and proton) densities are somewhat low, compared
to estimates of the explosion energy (only a few percent). The centroids
for the GeV to TeV emission seem to be shifted towards the western region
of the remnant where nonthermal X-ray emission is also
brightest~\citep{HelderVink2008, Maeda2009,PaperI}.

However, given the gamma-ray data published so far it was not possible
to unambiguously determine the particle population responsible for the
bulk of the emission, in particular to distinguish between gamma rays
produced through the bremsstrahlung and inverse
Compton (IC) leptonic processes
and the neutral pion decay hadronic process.
Lower-energy gamma rays (below 1~GeV) hold the key to
distinguishing between these scenarios, since a sharp low-energy
roll-over in the spectrum of hadronically-produced gamma rays is
expected~\citep{Stecker1970}. Continuous observations of Cas A with the \fermi-\lat
have provided us a better opportunity
to investigate the gamma-ray emission in the $\lesssim$ 1 GeV range.



The \lat is
a pair-conversion detector that operates between 20 \mev and $>300$
\gev. The telescope has been in routine scientific operation since 2008 August
4. With its wide field of view of 2.4~sr, the \lat observes the whole
sky every $\sim3$ hours. More details about the \lat instrument and its operation
can be found in \citet{LATPaper}. In addition, the data reduction process
and instrument response functions recently have been improved based on two
years of in-flight data \citep[so-called Pass7v6,][]{Pass7validation}.
According to the updated instrument performance, the point-spread function of
the \lat gives a 68\% containment angle of $<6\degree$ radius at 100~MeV and $<0\fdg3$
at $>10$~GeV for normal incidence photons in P7SOURCE class. The sensitivity of
the \lat~ for a point source with a power law photon spectrum of index 2 and a location similar to Cas A is $\sim 9\times10^{-9}$~\fluxunits for a $5\sigma$ detection above 100~MeV
after 44 months of sky survey.
Our analysis takes advantage of both the increase in data quantity and quality.

In this letter, we describe our analysis method in \secref{method},
present the \fermi results in \secref{results}, and then discuss
the gamma-ray emission mechanism of Cas A in \secref{discussion}.

\section{Analysis Method}\label{sec:method}

We analyzed \fermi-LAT observations of Cas A using data
collected from 2008 August 4 to 2012 April 18 (Mission elapsed time 239557565.63 -- 356436692.23, about 44 months of data).
The analysis was performed in the energy range 100~\mev-100~\gev using the \lat Science Tools\footnote{The
LAT Science Tools are distributed through the \textit{Fermi} Science Support
Center (FSSC, \url{http://fermi.gsfc.nasa.gov}).} as well as an independent tool \pointlike. In particular,
we used the maximum-likelihood fitting packages \pointlike to fit
the position and test for significant spatial extension of Cas A,
then with the updated localization result we used \gtlike to fit the spectrum of the source. Our analysis
procedure is very similar to that of the second LAT source catalog
\citep[2FGL, ][]{second_lat_catalog}.  When analyzing the data, we used
the P7SOURCE class event selection and P7\_V6 instrument response functions
\citep[IRFs,][]{Pass7validation}. In order to reduce contamination from gamma rays
produced in the Earth's limb, we excluded events with reconstructed zenith angle greater than 100\degree, and selected times when the rocking angle was less than 52\degree.

Emission produced by the interactions of cosmic rays with interstellar gas and radiation fields substantially contributes to the gamma-ray intensities measured by the LAT near the Galactic plane. We accounted for it using the standard diffuse model used in the 2FGL analysis. We also included the standard isotropic template accounting for the isotropic gamma-ray background and residual cosmic-ray contamination.\footnote{The diffuse model gal\_2yearp7v6\_v0.fits and isotropic template isotrop\_2year\_P76\_source\_v0.txt
can be obtained through the FSSC.} In addition, we modeled as background
sources all nearby 2FGL sources:
in \pointlike we used a circular region of interest (ROI) with a radius of $15\degree$
centered on Cas A; in \gtlike we used a square region of interest with a size of
$20\degree\times20\degree$ aligned with Galactic coordinates, using a spatial binning of $0\fdg125\times0\fdg125$. We adopt the same parameterizations as 2FGL for these sources, while left free the spectral parameters of 5 2FGL sources that were either nearby or had a significant residual when assuming the 2FGL values: 2FGL J2333.3+6237, 2FGL J2257.5+6222c, 2FGL J2239.8+5825, 2FGL J2238.4+5902, 2FGL J2229.0+6114. In addition, we added 4 sources not included in 2FGL which will be described in Section \secref{spatial}.

\section{Results}\label{sec:results}

\subsection{Spatial Analysis}\label{sec:spatial}

Because of the wide and energy-dependent point-spread function of
the LAT, nearby sources must be carefully modeled to avoid bias
during a spectral analysis.
Therefore, before analyzing Cas A,
we performed a dedicated search for nearby point-like sources
not included in the 2FGL catalog.
We did so by
adding sources in the background model at the positions of significant
residual test statistic \citep[\ts, which follows the same definition as that in][]{second_lat_catalog} until the residual $\ts<25$ within the entire \pointlike ROI. \tabref{new_sources} lists the
four significant new sources found in this study. We have not found any counterparts for the new sources yet.

\begin{deluxetable}{lccccc}
\tablecaption{New sources added to the ROI
\label{tab:new_sources}
}
\tablewidth{0pt}
\tablehead{
\colhead{Name} & \colhead{\ts} & \colhead{$l$} & \colhead{$b$} & \colhead{Flux}  & \colhead{Index}\\
\colhead{} & \colhead{} & \colhead{(deg.)} & \colhead{(deg.)} & \colhead{($10^{-8}$ \fluxunits)} & \colhead{}
}
\startdata
Source 1 & 35.0 & 120.10 &  1.41 & 1.96$\pm$0.53 & 2.24$\pm$0.10\\
Source 2 & 31.7 & 118.59 & $-1.14$ & 0.89$\pm$0.40 & 2.04$\pm$0.15 \\
Source 3 & 25.6 & 113.16 & $-0.28$ & 0.66$\pm$0.31 & 1.92$\pm$0.16 \\
Source 4 & 24.8 & 105.82 &  2.89 & 1.39$\pm$0.65 & 2.12$\pm$0.14 \\
\enddata
\tablecomments{The spectral and spatial parameters of the new sources
found in the region surrounding Cas A. $l$ and $b$ are the Galactic
longitude and latitude of the source and \ts is the significance of the
detection of the source (in the energy range from 100 \mev to 100 \gev).
The sources were modeled with a power-law spectral model and the flux
is computed from 100 \mev to 100 \gev.
}
\end{deluxetable}

\begin{figure}[htbp]
  \centering
   \ifdefined\bwfigures
     \includegraphics{countmap_bw.eps}\\
   \else
     \includegraphics{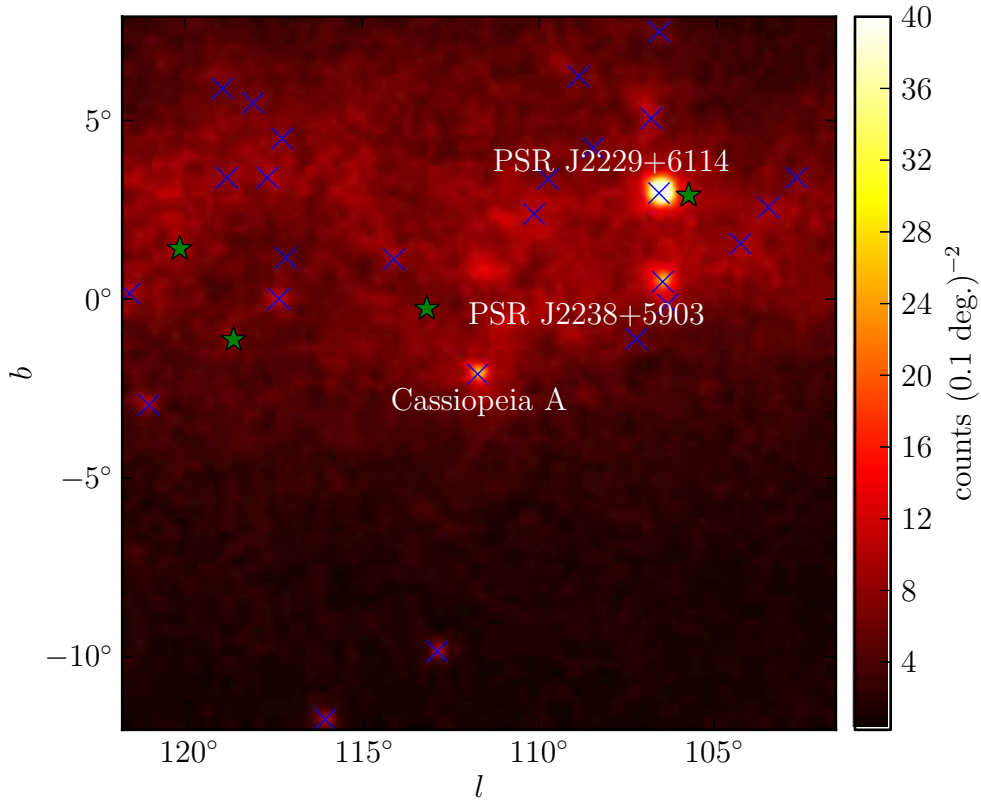}\\
   \fi
  \caption{\fermi-\lat count map of the
  region surrounding Cas A ($20\degree\times20\degree$)
  from 800 \mev to 100 \gev.
  This plot is smoothed by a Gaussian kernel of size 0\fdg1. Also shown are the 2FGL sources included in our background model (blue crosses) and the new sources we added in (green stars).}\label{fig:count_map}
\end{figure}

\figref{count_map} shows a count map above 800 MeV of the region surrounding Cas A.
The relatively bright source coincident with the
SNR Cas A has a \ts value of $\sim600$.  First, we
used \pointlike to fit the position of this source and test for any possible
spatial extension.  The best fit position of the source, in Galactic coordinates,
is $l,b=111\fdg74,-2\fdg12$,
with a statistical uncertainty of $0\fdg01$ (68\% containment).
To account for the systematic error in the position
of Cas A, we added $0\fdg005$ in quadrature as was adopted for the 2FGL analysis
\citep{second_lat_catalog}.

\begin{figure}
  \centering
  \ifdefined\bwfigures
     \includegraphics[width=0.6\textwidth]{localization_bw.eps}
   \else
     \includegraphics[width=0.6\textwidth]{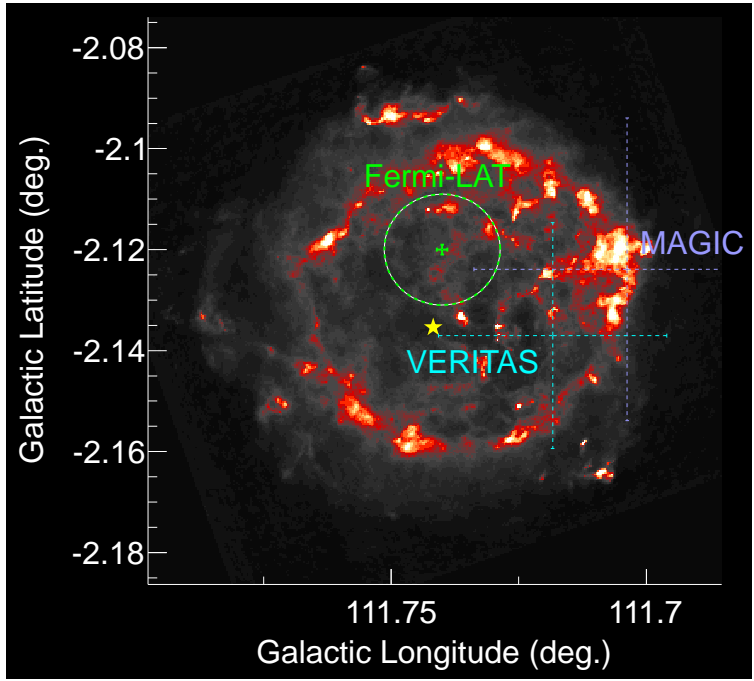}
   \fi
  \caption{\fermi-\lat best-fit localization of Cas A (shown as a green cross, also shown is the error ellipse at 68\% confidence level, calculated by adding statistical and systematic errors in quadrature), overlaid with VLA 20 cm radio map of the Cas A SNR \citep{Anderson1995}. The central compact object is shown as a yellow star. Also shown are best-fit positions obtained by MAGIC \citep{MAGIC:casA} and VERITAS \citep{VERITAS_CasA}.}\label{fig:localization}
\end{figure}

This location
is only $0\fdg02$ away from the central compact object (CCO)
\citep{PavlovLuna2009}, as shown in \figref{localization}. This confirms that the GeV source is most
likely the $\gamma$-ray counterpart of the Cas A SNR.  Following the method described in
\cite{Lande_2012_Extended_Sources}, we used a disk spatial model to fit the extension of Cas A.
We found that the emission was not
significantly spatially extended  ($\tsext=0.1$) and has an extension
upper limit of 0\fdg1~ at 95\% confidence level. Note that this upper limit is larger than the shell of Cas A.







\subsection{Spectral Analysis}

We performed a spectral analysis of Cas A in the energy range from 100~\mev to
100~\gev using \gtlike.
We first fit Cas A with a power-law
spectral model and found an integral flux of $(6.17\pm0.43_\stat)\times10^{-11}$
\efluxunits~ in the energy range from 100 \mev to 100 \gev
and a photon index of $\Gamma=1.80\pm0.04_\stat$.
The results are consistent with the previous analysis of \citet{PaperI}.


We then tested for a break in the spectrum of Cas A by fitting the spectrum with
a smoothly-broken power-law spectral model
\begin{equation}
    \frac{dN}{dE}=N_0\left(\frac{E}{E_0}\right)^{-\Gamma_1}\left(1+\left(\frac{E}{E_b}\right)^{\frac{\Gamma_2-\Gamma_1}{\beta}}\right)^{-\beta}.
\end{equation}
Here, $N_0$ is the prefactor;
$E_0$ is a fixed energy scale (taken to be 1 \gev);
$E_b$ is the break energy;
$\Gamma_1$ and $\Gamma_2$ are the photon indices before and after the break, respectively;
$\beta$ is a small, fixed parameter that describes the smoothness of
the transition at the break (taken to be 0.1).

We tested for the significance of
this spectral feature using a likelihood ratio test:
\begin{equation}
  \tsbreak = 2\log(\likelihood_\text{SBPL}/\likelihood_\text{PL})
\end{equation}
where \likelihood is the Poisson likelihood of observing the given data
assuming the best-fit model.  We obtained $\tsbreak=48.2$, indicating
that the break is significant.  The resulting spectral parameters are quoted in \tabref{sbpl_params}.

\begin{figure}[htbp]
  \centering
   \ifdefined\bwfigures
     \includegraphics{sed_new_bw.eps}
   \else
     \includegraphics{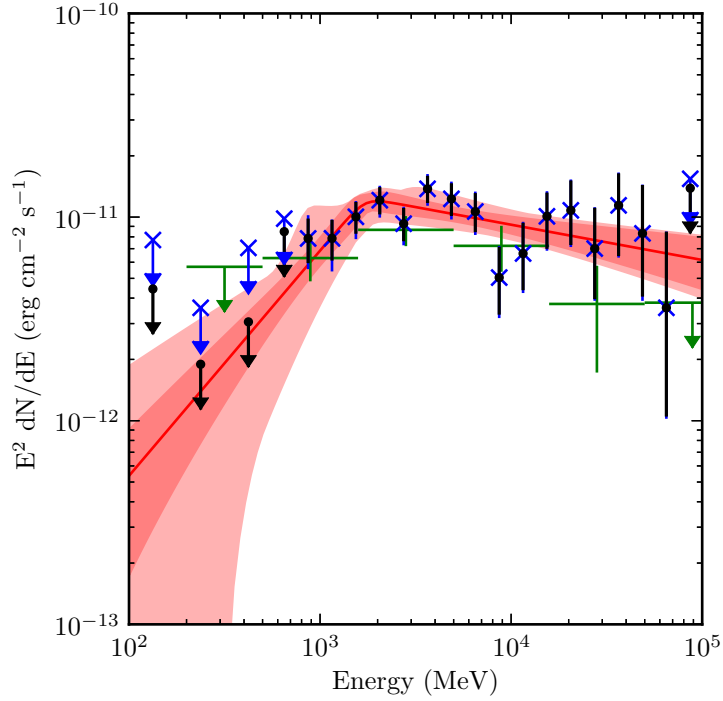}
   \fi
  \caption{The spectral energy distribution of Cas A.
  The black points include statistical error only
  and the blue cross points include both statistical and systematic errors added in quadrature.
  The black upper limits consider only statistical effects and are calculated at 95\% confidence level using a Bayesian method. We plot an upper limit instead of a data point when $\ts<4$. Blue upper limits have included systematic uncertainties.
  The red line is the best-fit spectral model assuming a smoothly-broken power law.
  The dark shaded region represents the statistical error on the spectral fit
  and the lightly shaded region represents the systematic and statistical errors added in quadrature.
  Also shown are the spectral points measured in Paper I (green points).
  }
  \label{fig:spectrum_sbpl}
\end{figure}

We then computed a spectral energy distribution (SED) in 8 bins per energy
decade by fitting the flux of Cas A independently in each energy bin
(the lowest 6 bins were combined into 3 bins). The SED of Cas A, along
with the all-energy spectral fit, is plotted in \figref{spectrum_sbpl}.
Statistical upper limits are shown in energy bins where \ts of the flux is less than 4. These upper limits are calculated at 95\% confidence level using a Bayesian method \citep[e.g.,][]{Helene1983319}.

\subsection{Systematic Errors}

We estimated the systematic errors on the spectrum of Cas A
due to uncertainty in our model of the Galactic diffuse emission and
due to uncertainty in our knowledge of the IRFs of the LAT.

To probe the uncertainties due to the modeling of Galactic diffuse
emission we use a series of alternative models \citep{dePalma2013}. These models differ from the standard one in the sense that \citeauthor{dePalma2013} 1) adopt different gamma-ray emissivities for the interstellar gas, different gas column densities, and use a different approach for incorporating spatially extended residuals; 2) vary a select number of important
input parameters of the model \citep{diffuse}:
the H~{\sc i} spin temperature, the cosmic-ray source distribution, and height of
the cosmic-ray propagation halo; 3) allow more freedom in the fit by separately
scaling components of the model in four Galactocentric rings. Although
these models do not span the complete uncertainty of the systematics
involved with Galactic diffuse emission modeling, they were selected to probe the most important systematic uncertainties.

\begin{figure}
  \centering
   \ifdefined\bwfigures
     \includegraphics{residual_compare2_bw.eps}\\
   \else
     \includegraphics{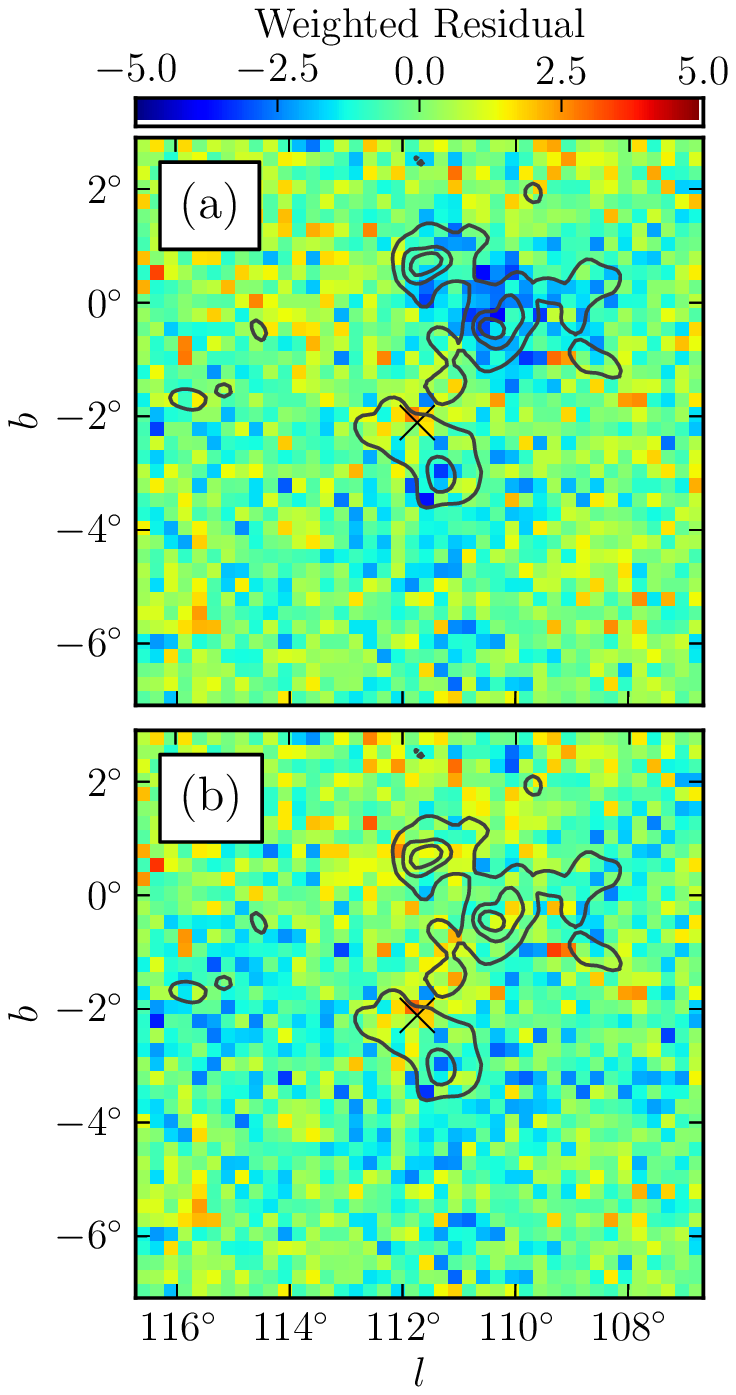}\\
   \fi
  \caption{Weighted residual count maps (unsmoothed) in the energy range 100 \mev to 100 \gev after fitting with (a)
  standard diffuse model and (b) one of the alternative diffuse models. The weighted residual $s$ is calculated as $s=(N_{\text{obs}}-N_{\text{mdl}})/\sqrt{N_{\text{mdl}}}$, where $N_{\text{obs}}$ and $N_{\text{mdl}}$ are observed count and model count, respectively. The location of Cas A is indicated by the black cross. The contours correspond to integrated intensity of the CO line and represent the
  column-density distribution of the molecular complex associated with NGC 7538 and Cas A
  \citep[this is the same CO intensity map of the Perseus arm with the same velocity range of integration as described in ][]{diffuseCasA}.
  The CO map was smoothed using
  a Gaussian kernel of 0\fdg5. Contours of 8, 29, and 50 K km s$^{-1}$
  are shown.}\label{fig:residual_compare}
\end{figure}

At low energy ($<1$~\gev),
our uncertainty in the modeling of the Galactic diffuse emission leads to significant uncertainty in the spectral analysis of Cas A, because the integrated intensity of the diffuse emission on the scale of the energy dependent point spread function of the \lat becomes comparable with the flux of the source. By examining the residual maps after fitting, we found that the standard diffuse model overshoots the data for a region $\sim2\degree$ from Cas A (\figref{residual_compare}), and this can lead to underestimated upper limits in the SED calculation.

This overestimation of diffuse count is most likely due to uncertainty in
modeling the gamma-ray emission from the molecular complex associated with NGC 7538 and Cas A in the Perseus arm
\citep[e.g.,][]{diffuseCasA}. The alternative diffuse models provide a qualitatively better fit of this region when the normalization of each Galactocentic ring was left free, since the increased degrees of freedom allow us to better scale the Galactic diffuse model for this specific region.
The improvement can be seen in
\figref{residual_compare} which shows a residual map with the standard
diffuse model and an improved residual map with one of the alternative diffuse models.

Even though there is significant systematic uncertainty in the spectral
model of Cas A at lower energies, \tsbreak was greater than 20 using all
of the alternative diffuse models and is therefore robust against this
systematic uncertainty.

We estimated the systematic error due to uncertainty in the IRFs using the
method described in \citet{Pass7validation}. Following
this method, we set the pivot in the bracketing IRFs at 2~\gev,
near the spectral peak in our SED.
Again, we found the spectral break to be robust against uncertainty in IRFs.

The systematic errors on the
estimated spectral parameters due to both systematic uncertainties are included
in \tabref{sbpl_params}.

\begin{deluxetable}{lccccc}
\tablecaption{Spectral Results for Cas A
\label{tab:sbpl_params}
}
\tablewidth{0pt}
\tablehead{
\colhead{Parameters} & \colhead{Value} & \colhead{$\Delta_\stat$} & \colhead{$\Delta_\text{sys,diffuse}$} & \colhead{$\Delta_\text{sys,IRFs}$} & \colhead{$\Delta_\text{sys}$}
}
\startdata
Energy flux ($10^{-11}\,\efluxunits$) &  4.69 & 0.38 & +0.03/-0.73 &   +0.45/-0.36 & +0.45/-0.81 \\
$\Gamma_1$                            &  0.89 & 0.29 & +0.46/-1.00 &   +0.32/-1.37 & +0.55/-1.70 \\
$\ebreak$ (\gev)                      &  1.72 & 0.40 & +0.62/-0.17 &  +1.20/-0.87 & +1.35/-0.89 \\
$\Gamma_2$                            & 2.17 & 0.09 & +0.06/-0.02 &   +0.08/-0.05 & +0.10/-0.05\\
\enddata
\tablecomments{Spectral fit of Cas A assuming a smoothly-broken power-law spectral model.
Energy flux is quoted from 100 \mev to 100 \gev. $\Delta_\stat$ is the
statistical error; $\Delta_\text{sys,diffuse}$ is the estimated
systematic error due to uncertainties in modeling the Galactic diffuse emission; $\Delta_\text{sys,IRFs}$ is the estimated systematic error
due to uncertainty in our knowledge of the IRFs of the LAT. $\Delta_{\text{sys}}$ is derived by adding the two components of systematic errors in quadrature.}
\end{deluxetable}

\section{Discussion}
\label{sec:discussion}

In \figref{spectrum_sbpl}, the new spectral data points measured
with the \emph{Fermi}-LAT are overlaid with those from Paper~I.
The newly-measured spectrum is consistent with the previous result,
except that most of the new data points lie slightly above the old measurement. This is likely due to the changed event classifications and improved IRFs of the \lat as well as updated background models.
In Paper~I, we argued that the GeV--TeV gamma rays
detected from Cas~A can be interpreted in terms of either a leptonic or a hadronic model.
In these models, cosmic-ray electrons and protons (and ions) are accelerated in Cas~A and produce the gamma-ray emission.
In what follows, we revisit the gamma-ray emission models and then discuss
the new LAT spectrum.

The synchrotron X-ray filaments found at the locations of outer shock waves
indicate efficient acceleration of cosmic-ray electrons at the forward shocks
\citep{Hughes2000,Gotthelf2001,Vink2003,Bamba2005,PatnaudeFesen2009}.
Moreover, X-ray studies with \emph{Chandra} suggest that
electron acceleration to multi-TeV energies also takes place
 at the reverse shock propagating inside the supernova ejecta \citep{Uchiyama2008,HelderVink2008}.
The detections of TeV gamma rays with HEGRA~\citep{HEGRA:casA},
MAGIC~\citep{MAGIC:casA} and VERITAS~\citep{VERITAS_CasA},
established the acceleration of multi-TeV particles in the remnant.
Because of the small radius of $2.5\arcmin$ of Cas~A, these experiments lacked
the angular  resolution to determine the spatial distribution of the gamma rays
and the sites of particle acceleration.

It is widely considered that diffusive shock acceleration
\citep[DSA: see e.g.,][for a review]{MalkovDrury} operating at the forward shocks
is responsible for the energization of the cosmic-ray particles.
Most DSA models, which provide predictions of gamma-ray spectra of SNRs, focus on
the acceleration at the forward shock \citep[e.g.,][]{Ellison10,Morlino12}.
Recently, newly-developed non-linear DSA models have included the effects of
acceleration of particles at reverse shocks and their subsequent transport \citep{ZP12}.
\citet{Zira13} have demonstrated that about 50\% of the gamma-ray flux at 1 TeV
from Cas~A can be contributed by the reverse-shocked medium.
Although the nonthermal X-ray filaments and knots in the reverse-shock
region are interesting sites of particle acceleration \citep{Uchiyama2008}, we assume that
the gamma-ray emission comes predominantly from the forward shock
region. Note that our discussion on  leptonic versus hadronic emission would not be greatly affected by
this assumption, because we allow for parameter space that is relevant also for the reverse-shocked regions.


The gamma-ray emission models are constrained by the gas and radiation density
and by the magnetic field in the gamma-ray production region.
We assume the simplest model where cosmic rays are distributed
uniformly in the shell of the remnant.
The fluxes of bremsstrahlung and $\pi^0$-decay gamma-ray emission scale
linearly with the average gas density ($\propto \bar{n}$).
Likewise the IC flux is proportional to the radiation energy density
($\propto U_{\rm ph}$) as long as IC scattering is in the Thomson regime.
The synchrotron flux scales as $\propto B^{(s+1)/2}$ for a fixed density of
electrons with a power-law index of $s$.
The magnetic field only indirectly affects the gamma-ray flux by
determining the amount of relativistic electrons that are required to produce
the observed synchrotron radio emission.
This  in turn can be used to calculate the bremsstrahlung
and IC fluxes. Therefore the gamma-ray flux constrains the magnetic
field in the shell~\citep{Cowsik}.

The outer shock waves are currently propagating into a dense circumstellar wind.
The density behind the blastwave is estimated as
 $n_{\rm H} \simeq 10\ {\rm cm^{-3}}$ from the measured hydrodynamical
 quantities such as shock velocities \citep{LamingHwang2003}.
  The radiation field for IC scattering is
dominated by far infrared (FIR) emission from the shock-heated ejecta,
characterized by a temperature of 100~K and an energy density of
$\sim 2\ \rm{eV}\ \rm{cm}^{-3}$ \citep{Mezger1986}.
Using the gas and infrared densities, which are well constrained from
the multiwavelength data, it was shown in Paper I that
 bremsstrahlung by relativistic electrons dominates the leptonic component
below $\sim 1$ GeV, and IC/FIR becomes comparable to bremsstrahlung above 10 GeV,
for the assumed electron acceleration spectrum $Q_{\rm e}(E) \propto
E^{-2.34} \exp( -E/E_{\rm m} )$ with $E_{\rm m} = 40$ TeV \citep{Vink2003}.
The power-law index was set
to match the radio-infrared spectral index of $\alpha = 0.67$
\citep{Rho}, since both the GeV gamma-ray emission and the radio
synchrotron emission sample similar electron energies.
We note that the IC scattering of FIR  exceeds IC of cosmic microwave background
by a factor of $\sim 3$ at 10 GeV.

\begin{figure}[htbp]
  \centering
   \ifdefined\bwfigures
     \includegraphics{model8bpd_final_large_bw.eps}
   \else
     \includegraphics{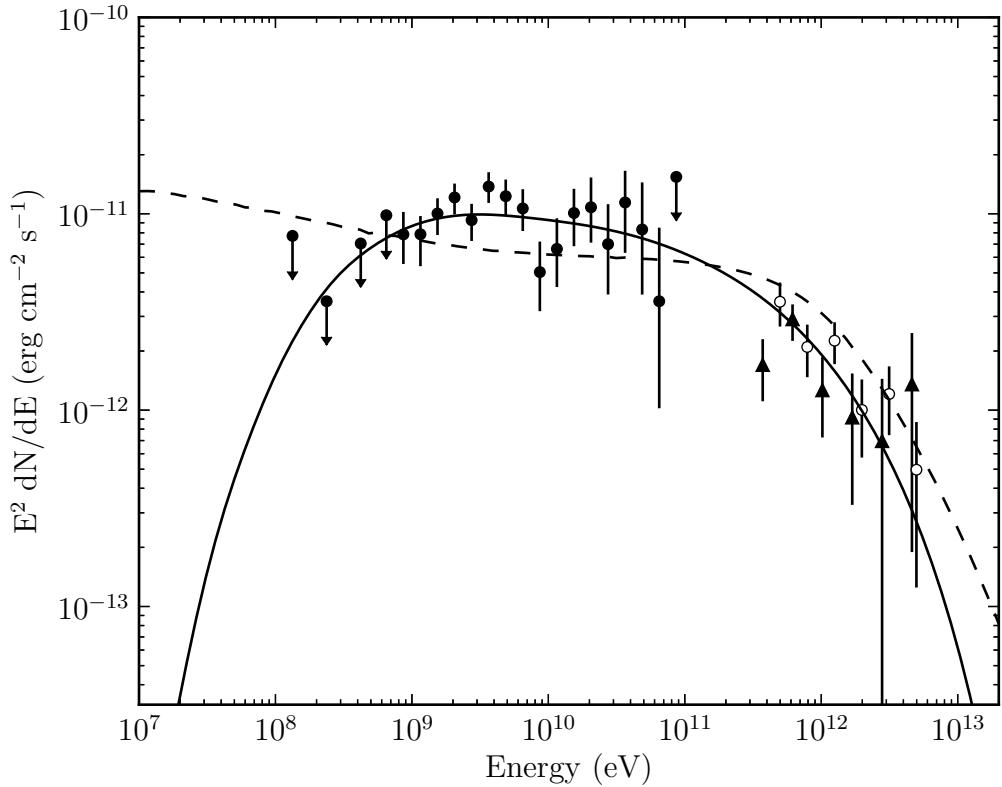}
   \fi
   \caption{Gamma-ray spectrum of Cas~A together with the
   emission models. The \emph{Fermi}, MAGIC, and VERITAS points are plotted as
   filled circles, triangles and open circles, respectively
   \citep[][]{MAGIC:casA,VERITAS_CasA}. The \emph{Fermi} spectral points include both statistical and systematic errors.
   The curves show a leptonic model for $B=0.12\ {\rm mG}$ (dashed line) and
   the hadronic model from Paper~I with its normalization
   increased by $27\%$ (solid line). }
 \label{fig:SED_models}
\end{figure}

Figure~\ref{fig:SED_models} compares the leptonic model presented in Paper I
with our new LAT measurement. The magnetic field
$B=0.1\ {\rm mG}$ used in the leptonic model  is consistent with
$B=0.08\mbox{--}0.16\ {\rm mG}$ estimated by \citet{Vink2003}
who interpreted the width of a synchrotron X-ray filament as
the  synchrotron cooling length.
The field is
somewhat lower than $B \simeq 0.3\ {\rm mG}$ estimated by \citet{Parizot2006}
who took into account a projection effect.
Unlike the TeV band where the electrons responsible for the
gamma-ray emission suffer from severe synchrotron losses, the gamma-ray
spectral shape near 1 GeV does not depend on the magnetic field.
This can be seen, for example, in \citet{Araya10} who employed
different magnetic field strengths (by a factor of 6) between two radiation zones.

Also shown in Figure~\ref{fig:SED_models} is the hadronic model presented
in Paper~I. To achieve a better match with the new measurement,
the normalization of the model spectrum is increased by 27\% from Paper~I.
The model was calculated for
a proton spectrum of $Q_{\rm p}(p)\propto p^{-2.1} \exp (-p / p_{\rm m} )$
 with an exponential cutoff at $cp_{\rm m} = 10$ TeV, where
 $p$ denotes momentum of accelerated protons.
The total proton content amounts to $W_p (>10\ {\rm MeV}\,c^{-1})
\simeq 4\times 10^{49}\ {\rm erg}$, which is
 less than 2\% of the estimated explosion kinetic energy of $E_{\rm sn} =
2\times 10^{51}\ {\rm erg}$ \citep{LamingHwang2003,HwangLaming2003}\footnote{
The shocked ejecta gas can contribute to the gamma-ray emission.
 The baryon density in the shocked ejecta is similar to that  in the forward shock region.
Therefore,
the total proton content estimated here can be interpreted roughly as a sum of the
cosmic-ray contents in the forward shock region and that in the reverse-shocked
ejecta. }.

Paper~I already showed that
the leptonic model cannot fit  the turnover well at low energies because
the bremsstrahlung component that is dominant over IC below 1 GeV
has a steep spectrum.
Note that the spectral shape of the bremsstrahlung component copies
the electron spectrum with spectral index $s=2.34$, which in turn is determined from the radio-infrared spectral index of $\alpha = 0.67$ \citep{Rho}. If we use a steeper power law for the electron energy distribution based on
a global spectral index of $\alpha = 0.77$ in the radio wavelengths \citep{Baars} or a spectral shape with curvature that reproduces the hardening ($\alpha = 0.77 \rightarrow 0.67$) in the integrated spectrum, the discrepancies
between the bremsstrahlung model and the \emph{Fermi}-LAT data become even larger.
\citet{Araya10}, who reported the results of \emph{Fermi}-LAT analysis of Cas~A independently,  also showed that the electron bremsstrahlung with such a steep electron index could not explain the \emph{Fermi}-LAT spectrum.
However,
 uncertainties in the Galactic diffuse emission at low energies prevented a definitive
 conclusion regarding the inconsistency between the bremsstrahlung model
 and the gamma-ray data.
In this paper, a more detailed investigation of these uncertainties at low energy
now  confirms the hadronic origin of the GeV $\gamma$-ray emission from Cas~A.
The new LAT spectrum can be described by a broken power law
with a second power-law index of
$\Gamma_2 = 2.17 \pm 0.09$.
A comparison between  the LAT spectrum and  the TeV $\gamma$-ray spectra suggests
that additional steepening between the LAT and the TeV bands is necessary.
Indeed,
the TeV $\gamma$-ray spectra measured with HEGRA, MAGIC, and VERITAS
are consistent with a power law with a photon index of
$\Gamma_{\rm TeV} = 2.5\pm 0.4_{\rm stat} \pm 0.1_{\rm sys}$,
$\Gamma_{\rm TeV} = 2.3\pm 0.2_{\rm stat} \pm 0.2_{\rm sys}$, and
$\Gamma_{\rm TeV}  = 2.61\pm 0.24_{\rm stat} \pm 0.2_{\rm sys}$, respectively,
which are somewhat steeper than the second index $\Gamma_2 = 2.17 \pm 0.09$ of the
LAT spectrum.
However, given the relatively large statistical uncertainties of the TeV $\gamma$-ray fluxes,
we refrain from solidifying the presence of the cutoff.
If confirmed, efficient acceleration of particles to PeV energies in Cas~A is questioned.

The \emph{Fermi}-LAT results on two historical SNRs, Tycho's SNR \citep{Tycho_LAT12} and Cas~A, support hadronic scenarios for these objects. Tycho's SNR is the remnant of a Type Ia supernova, while Cas~A is that of a core-collapse SN (specifically Type IIb). This indicates that both Type Ia and core-collapse SNRs can convert a substantial fraction of
their kinetic expansion energies into cosmic-ray energies, and makes SNRs energetically favorable  candidates for the origin of Galactic cosmic rays. Recently, direct spectral signatures of
the $\pi^0$-decay emission have been found in two middle-aged SNRs interacting with molecular clouds: W44 and IC 443 \citep{PionSNR,AGILE_W44}. Although spectroscopic evidence for
the $\pi^0$-decay emission from Cas A is not as strong as these two cases, our results presented in this paper demonstrate the importance of the gamma-ray measurements of SNRs below 1 GeV.

The \fermi-LAT Collaboration acknowledges generous ongoing support
from a number of agencies and institutes that have supported both the
development and the operation of the LAT as well as scientific data analysis.
These include the National Aeronautics and Space Administration and the
Department of Energy in the United States, the Commissariat \`a l'Energie Atomique
and the Centre National de la Recherche Scientifique / Institut National de Physique
Nucl\'eaire et de Physique des Particules in France, the Agenzia Spaziale Italiana
and the Istituto Nazionale di Fisica Nucleare in Italy, the Ministry of Education,
Culture, Sports, Science and Technology (MEXT), High Energy Accelerator Research
Organization (KEK) and Japan Aerospace Exploration Agency (JAXA) in Japan, and
the K.~A.~Wallenberg Foundation, the Swedish Research Council and the
Swedish National Space Board in Sweden.

Additional support for science analysis during the operations phase is gratefully
acknowledged from the Istituto Nazionale di Astrofisica in Italy and the Centre National d'\'Etudes Spatiales in France.

\bibliography{casa_2nd_lat}

\end{document}